\documentclass[a4paper,11pt]{article}
\usepackage{pos}
\usepackage{cleveref}

\title{NUTRIG: Towards an Autonomous Radio Trigger for GRAND}
 \ShortTitle{NUTRIG}

\author*[a]{Pablo Correa}

\affiliation[a]{Sorbonne Université, Université Paris Diderot, Sorbonne Paris Cité, CNRS, Laboratoire de Physique Nucléaire et de Hautes Energies (LPNHE),\\
  4 place Jussieu, F-75252, Paris Cedex 5, France}

\onbehalf{for the GRAND collaboration} 

\emailAdd{pablo.correa@lpnhe.in2p3.fr}

\abstract{One of the major challenges for large-scale radio surface arrays, such as the Giant Radio Array for Neutrino Detection (GRAND), is the requirement of an autonomous online trigger for radio signals induced by extensive air showers. The NUTRIG project lays the foundations for the development of a pure, efficient, and scalable trigger in the context of GRAND. For this purpose, a GRAND prototype setup of four detection units has been deployed at Nançay, France, which currently serves as the main testing facility for the deployment of this autonomous trigger. This work provides a detailed description of the GRAND@Nançay setup, and a first analysis of background data gathered on site. Initial tests of signal recovery in laboratory conditions are also presented. Finally, near-future plans are outlined to scale NUTRIG to larger pathfinder arrays such as GRANDProto300.}

\ConferenceLogo{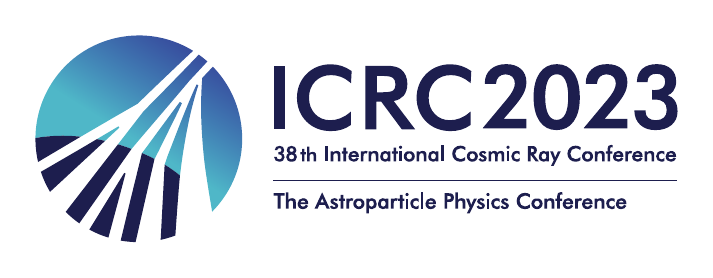}

\FullConference{%
38th International Cosmic Ray Conference (ICRC2023)\\
  26 July - 3 August, 2023\\
  Nagoya, Japan}

\begin{document}
\maketitle

\section{Introduction}

The Giant Radio Array for Neutrino Detection (GRAND) \cite{Alvarez_2020,Torres_2023} is a proposed surface array that primarily targets the detection of ultra-high-energy (UHE; $>$100 PeV) neutrinos. In particular, Earth-skimming UHE tau neutrinos can induce very-inclined extensive air showers with zenith angles near $90^\circ$. It is the transient ($\lesssim$100 ns) radio emission (coherent between $\sim$10 MHz up to several hundreds of MHz) produced by these very-inclined air showers via geomagnetic and Askaryan effects that GRAND aims to detect.



Due to the low expected flux of UHE neutrinos, GRAND is planned to instrument a total surface area of 200,000 $\mathrm{km^2}$ with detection units (DUs). A DU consists of a three-armed butterfly antenna and its front-end electronics. The inter-DU spacing will be relatively sparse (of the order of 1 km) since the radio footprint of very-inclined air showers is typically of the order of several $10~ \mathrm{km^2}$ \cite{Aab_2018}. More concretely, in its final stage, GRAND is envisaged to consist of 20 complementary arrays of 10,000 DUs each, spread across the globe for maximal sky coverage. Currently, the GRANDProto300 (China) and GRAND@Auger (Argentina) prototype arrays serve as pathfinders for GRAND10k arrays in the Northern and Southern Hemispheres, respectively. 




A cost-efficient deployment of gigantic radio arrays such as GRAND strongly relies on the development of an autonomous online radio trigger for air-shower signals, without the usage of external particle detectors. Previous efforts by CODALEMA \cite{Lautridou_2012}, AERA \cite{Asch_2008,Schmidt_2011}, and TREND \cite{Charrier_2018} have investigated the feasibility of utilizing an autonomous radio trigger at the antenna level. Although some positive results were achieved using this autonomous technique, the data-acquisition (DAQ) systems of these projects were not designed to handle the high trigger rates due to transient radio-frequency interference (RFI) experienced at their experimental sites. 

The NUTRIG project presented in this work lays the foundation for the development of an autonomous radio trigger for GRAND. It is a joint effort between KIT (Germany), and the GRAND Paris group (LPNHE and IAP, France). Both the principle and strategy of NUTRIG are outlined in Section~\ref{sec:NUTRIG}, as well as a description of the dedicated GRAND@Nançay prototype (France). A preliminary analysis of on-site (transient) RFI at GRAND@Nançay is also presented in Section~\ref{sec:analysis}. Finally, an outlook of short-term NUTRIG plans is given in Section~\ref{sec:outlook}.



\section{NUTRIG}
\label{sec:NUTRIG}

\subsection{Principle}

The NUTRIG principle is based upon the realization of three major objectives. These objectives, which are interconnected and being developed in parallel, are outlined below:
\begin{itemize}
    \item \textbf{Radio-emission model}. A detailed modeling of the radio emission of very-inclined (neutrino-induced) air showers is required in order to exploit its features at the trigger level. Efforts to model this very-inclined air-shower emission have previously been performed in the 30--80 MHz frequency band \cite{Huege_2019,Schluter_2023}. However, our signal model needs to be expanded to the 30--230 MHz range in which GRAND operates. Moreover, the signal model needs to be adapted to the specific location of a GRAND (prototype) array, since it will, amongst other aspects, depend on the local geomagnetic field, atmospheric conditions, and the altitude of the site.
    
    \item \textbf{First-level trigger (FLT)}. At the DU level, an air-shower signal will manifest itself as a transient voltage pulse in the recorded time trace. Such an air-shower pulse is expected to have specific characteristics (e.g.~time structure and signal polarization) which will be exploited by the FLT. Doing so, we aim at improving the background-rejection efficiency compared to a signal-over-threshold trigger, where the threshold is a few times above the stationary-noise level. Furthermore, to handle the limitations posed by the DAQ and the data-communication bandwidth, the target rate of the FLT can be no more than 100 Hz. For the same signal-selection efficiency as the threshold trigger currently used in GRAND prototypes, which saturates the DAQ at a trigger rate of 1 kHz, the FLT would therefore yield up to a factor 10 improvement in signal purity.

    \item \textbf{Second-level trigger (SLT)}. At the array level, the SLT will use data from the DUs where the FLT condition has been satisfied. Using the dedicated radio-emission model of air showers described above, we aim to significantly increase the signal purity by reducing the contamination of anthropogenic RFI and thermal noise. The quantity of information passed on from the FLT to the SLT will depend on the available communication bandwidth between the DUs and the DAQ system. Data of events that fulfill the SLT requirements will be recorded on disk for further offline analysis (see also \cite{Mitra_2023}).
\end{itemize}



\subsection{GRAND@Nançay}
\label{sec:nancay}

GRAND@Nançay is a prototype array primarily dedicated to the NUTRIG project. It consists of four GRAND-prototype DUs, and it is located at the Nançay Radio Observatory, deep in the forest of the French Sologne region. This protected radio-quiet environment was previously home to the CODALEMA experiment, which managed to successfully detect radio emission of cosmic-ray air showers \cite{Ardouin_2006}. Although GRAND@Nançay is not designed to identify air-shower signals, its relatively radio-quiet location provides an excellent setting for the on-site development of NUTRIG, and the FLT in particular.

\begin{figure}
    \centering
    \includegraphics[width=\textwidth]{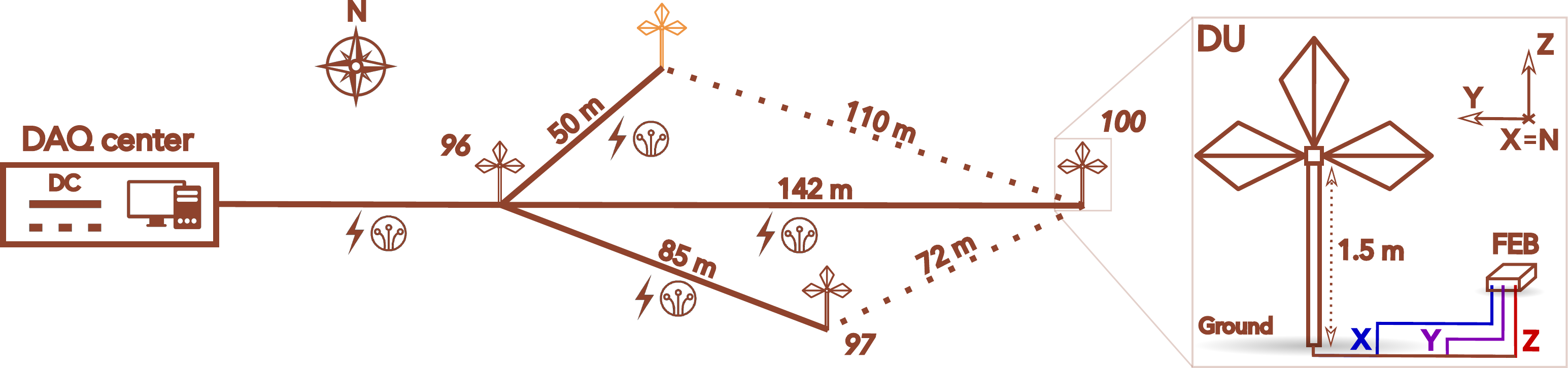}
    \caption{Schematic of the GRAND@Nançay prototype, which is dedicated to the development of NUTRIG. The setup currently consists of three active DUs (labeled 96, 97, and 100) with a fourth DU (indicated in light orange) planned to be deployed over the Summer of 2023. The DAQ center contains a DC power supply as well as a computer for data readout and storage. Data transfer and power supply to the four DUs is performed via optical fibers and coaxial cables, respectively (solid brown lines). The inset on the right shows a more detailed illustration of a single DU, where each of the three butterfly-antenna arms is connected to its corresponding channel on the FEB. Note that the antenna arm oriented along the $X$ axis is not shown for illustrative purposes.}
    \label{fig:nancay}
\end{figure}

The GRAND@Nançay setup and layout are illustrated in Fig.~\ref{fig:nancay}. Each of the four prototype DUs operate using a butterfly-antenna design with three orthogonal arms, which are oriented along the East-West axis ($X$), the North-South axis ($Y$), and upwards ($Z$). For each of the antenna arms, a captured signal is first amplified with a low-noise amplifier (LNA) located inside the antenna nut. Subsequently, the amplified signal is sent via coaxial cables from the LNA to the front-end electronics board (FEB) placed at the foot of the antenna.

Inside the FEB, both a band-pass filter of 30--230 MHz and a band-stop filter\footnote{Note that these FM filters are not present in the GRANDProto300 and GRAND@Auger prototypes.} in the FM band (87--108 MHz) are applied to the signal. After that, the signal passes through a variable gain amplifier (VGA; with an adjustable gain up to $23.5$ dB) before arriving at a 14-bit analog-to-digital converter (ADC). This ADC has four channels, three of which are labeled $X$, $Y$, and $Z$ and connected to the corresponding antenna arms, while the fourth serves as a floating channel. The signal is digitized by the ADC at a rate of 500 Msamples/s, and it is finally processed by the systems-on-chip (SoC) consisting of one field-programmable gate array (FPGA) and four central processing units (CPUs), which are used to implement the trigger logic and to build events at the DU level. In addition, four notch filters are implemented at the FPGA level.

DU events are transmitted to a computer in the central data-acquisition (DAQ) center via optical fibers\footnote{In GRANDProto300 and GRAND@Auger, the DU-DAQ communication is performed via a WiFi connection. However, this is not possible for GRAND@Nançay, where we would otherwise pollute the other experiments at the Nançay Radio Observatory. The optical fibers and FM band-stop filter are the only modifications to the FEB design deployed at GRANDProto300 and GRAND@Auger.}. The DAQ computer not only allows us to store data, but also to configure various FEB components, such as the ADC, VGA, and trigger logic. In addition, the central DAQ center also hosts an adjustable DC-power supply, which powers the FEBs via coax cables. Note that the LNA is powered via the FEB through the cable connected to the $Z$ channel.

\subsection{Trigger-Implementation Strategy}

The testing and optimization of the FLT and SLT algorithms requires a detailed characterization of both background RFI and expected air-shower signals at the DU level. For the description of the background, we use experimental data taken at the GRAND-prototype sites (GRAND@Nançay in particular), while air-shower signals are described using simulations. These simulations are currently based on the CoREAS and ZHaireS frameworks \cite{Huege_2013,Alvarez-Muniz_2011}, and also contain a complete description of the antenna response and front-end electronics chain. As such, we simulate the ADC voltage expected for air-shower pulses, which are superposed to recorded background data to obtain realistic signal-plus-background voltage traces. Such traces are currently being used to investigate different implementations of the FLT, such as machine-learning (see \cite{LeCoz_2023} for more details), template-fitting, and wavelet-analysis techniques.

Whereas the SLT algorithm will be implemented at the DAQ level of a GRAND array, the FLT algorithm will be encoded on the SoC of a FEB. Next, a first test of the online performances of this FLT algorithm will be performed with dedicated test bench at the LPNHE in Paris. Using a custom-wave function generator, simulated air-shower pulses are fed to the FEB, such that candidate FLT algorithms can be tested under controlled laboratory conditions.

\begin{figure}[t]
    \centering
    \includegraphics[width=0.495\textwidth]{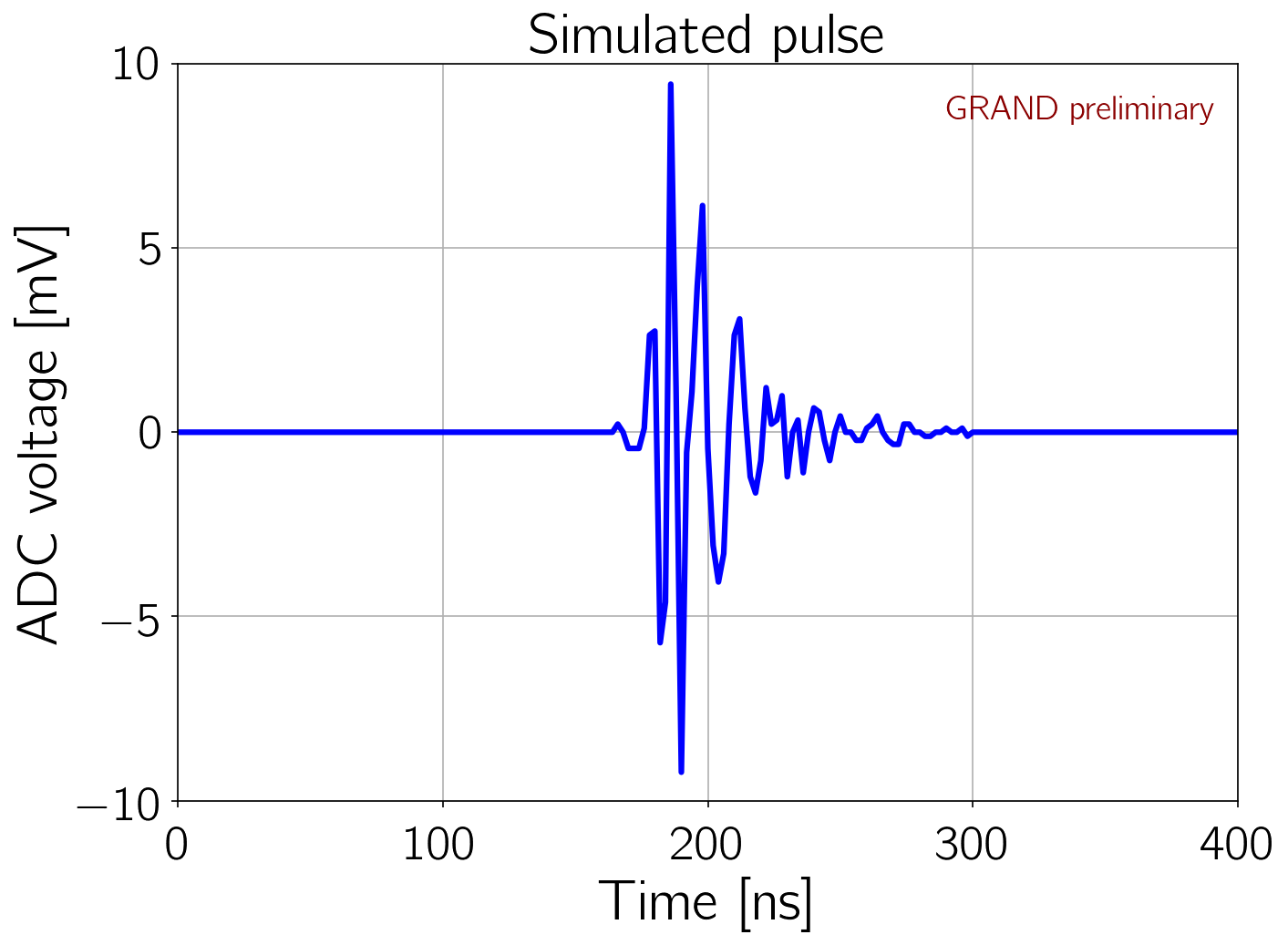}
    \includegraphics[width=0.495\textwidth]{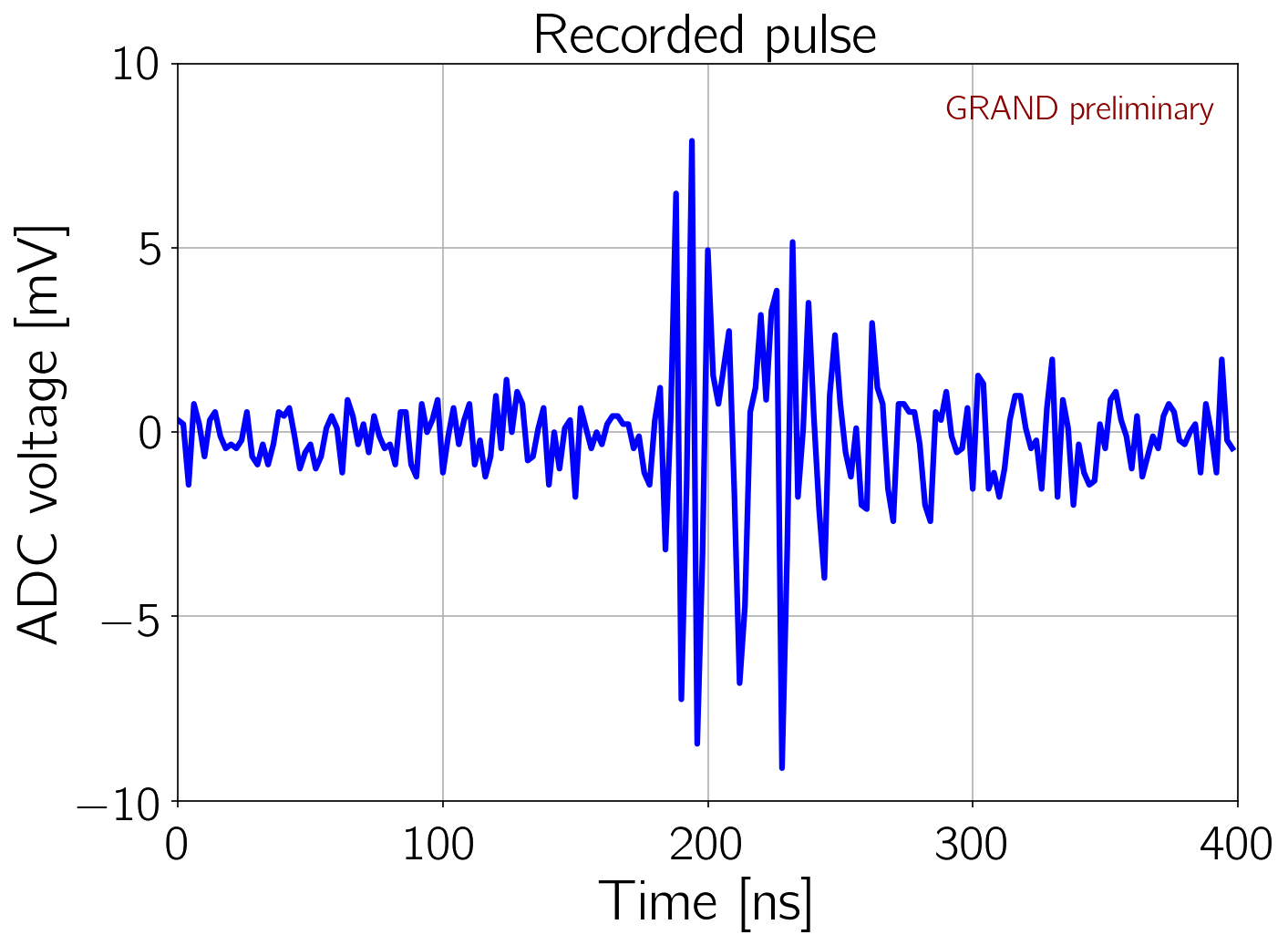}
    \caption{\textit{Left}: Time trace of a simulated air-shower pulse that has been processed through the entire front-end-electronics chain up to the ADC. Electronic noise is not included. \textit{Right}: Time trace of the same air-shower pulse recorded by an FEB at the LPNHE test bench in Paris, after being simulated at the antenna-output level and injected to the FEB by a custom-wave function generator.}
    \label{fig:generator}
\end{figure}

To validate our test-bench setup at the LPNHE, we use a simulated air-shower pulse for a DU located at the GRANDProto300 site. The electric field of the air shower at the DU position is generated with ZHaireS\footnote{For our test, we use a $10^{18}$ eV proton primary with zenith and azimuth angles of $85.5^\circ$ and $253^\circ$, respectively.}, convoluted with the antenna response function for the $X$ arm, and further processed through the front-end electronics (see \cite{LeCoz_2023} for more details on the complete simulation chain). First, we process the air-shower pulse through the complete chain up to the ADC, as shown in the left panel of Fig.~\ref{fig:generator}. Note that electronic noise produced by the front-end components is not yet included in the simulation. Next, we compute the pulse shape for the same signal at the output of the antenna nut. This intermediate pulse shape is then injected with our custom-wave function generator to an FEB, with which we record the measured ADC voltage, as shown in the right panel of Fig.~\ref{fig:generator}. We find that we can successfully measure the injected air-shower pulse produced with our custom-wave function generator. In addition, we find that our current simulation of the electronic chain after the antenna-nut output provides a reasonable description of the real electronics. 



Subsequently, four FEBs with the updated FLT algorithms will be deployed at GRAND@Nan-çay to be tested in the field. This will also allow us to determine which parameters to pass on from the FLT to the SLT algorithm, which will also utilize the dedicated radio-emission model for very-inclined air showers. Finally, the implementation of the complete FLT and SLT setup will be tested at GRANDProto300, which currently consists of 13 active DUs with 70 more ready to be deployed in the coming months. When completed, it will consist of 300 DUs which will allow to test the scalability of our trigger functionality. In any case, the GRANDProto300 pathfinder will be pivotal to ensure the scalability of NUTRIG to GRAND10k arrays.


\section{Analysis of GRAND@Nançay Background Data}
\label{sec:analysis}

Between 26--27 June 2023, we used the setup described in Section \ref{sec:nancay} to perform a preliminary characterization of the RFI background at GRAND@Nançay. The data for this analysis was taken between roughly 00:49 and 09:33 local time (UTC+2). Every ten seconds, each DU in the setup was forced to trigger the acquisition of data. This data-taking mode has the benefit that it allows us to test the stability of our setup during relatively long data-taking runs.

\begin{figure}
    \centering
    \includegraphics[width=0.6\textwidth]{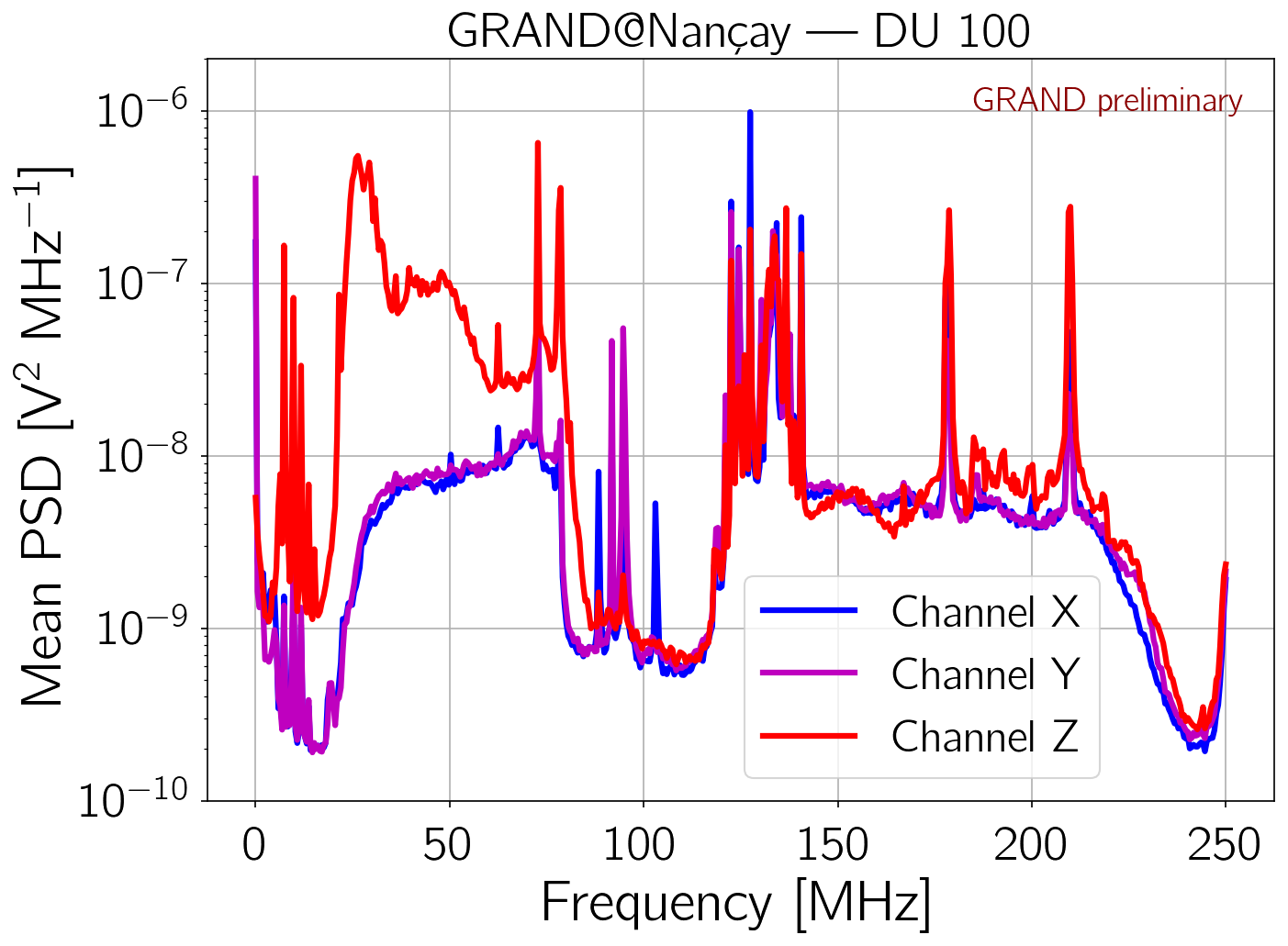}
    \caption{Mean PSD observed by DU 100 of the GRAND@Nançay setup. The spectra of channels $X$, $Y$, and $Z$ are shown in blue, magenta, and red, respectively. The various features of the PSD are discussed in the text.}
    \label{fig:fft}
\end{figure}

Figure \ref{fig:fft} shows the typical mean power spectral density (PSD) observed at GRAND@Nançay, in this case recorded by DU 100. The average was taken over the entire data sample. This PSD was obtained using all collected data and averaging out the spectrum over time. A first observation is that the PSD of channel $Z$ is up to a factor 50 higher than the PSD of channels $X$ and $Y$, most notably at frequencies below $\sim$70 MHz. This could be a consequence of the antenna transfer function, which is different for the $Z$ arm than for the symmetric $X$ and $Y$ arm. In addition, the LNA currently used at GRAND@Nançay has a different design for channel $Z$ compared to the other two channels, because the corresponding antenna arm is monopolar. An updated LNA is currently being tested at GRANDProto300 \cite{Ma_2023}.

In the features of the PSD in Fig.~\ref{fig:fft}, we can clearly observe the effect of both the 30--230 MHz band-pass filter and FM band-stop filter. However, short waves below 30 MHz and FM lines from nearby radio stations are still detected despite the filters. The two spectral lines at 72 MHz and 79 MHz are well-known radio emitters of the Nançay Radio Observatory, while the 120--140 MHz band is used for aeronautic communications. Finally, the peaks detected at 178 MHz and 210 MHz correspond to digital audio broadcasting lines. Nevertheless, we note that the notch filter implemented on the FPGA of the FEB will allow us to filter out constant wave emitters. 

\begin{figure}
    \centering
    \includegraphics[width=0.6\textwidth]{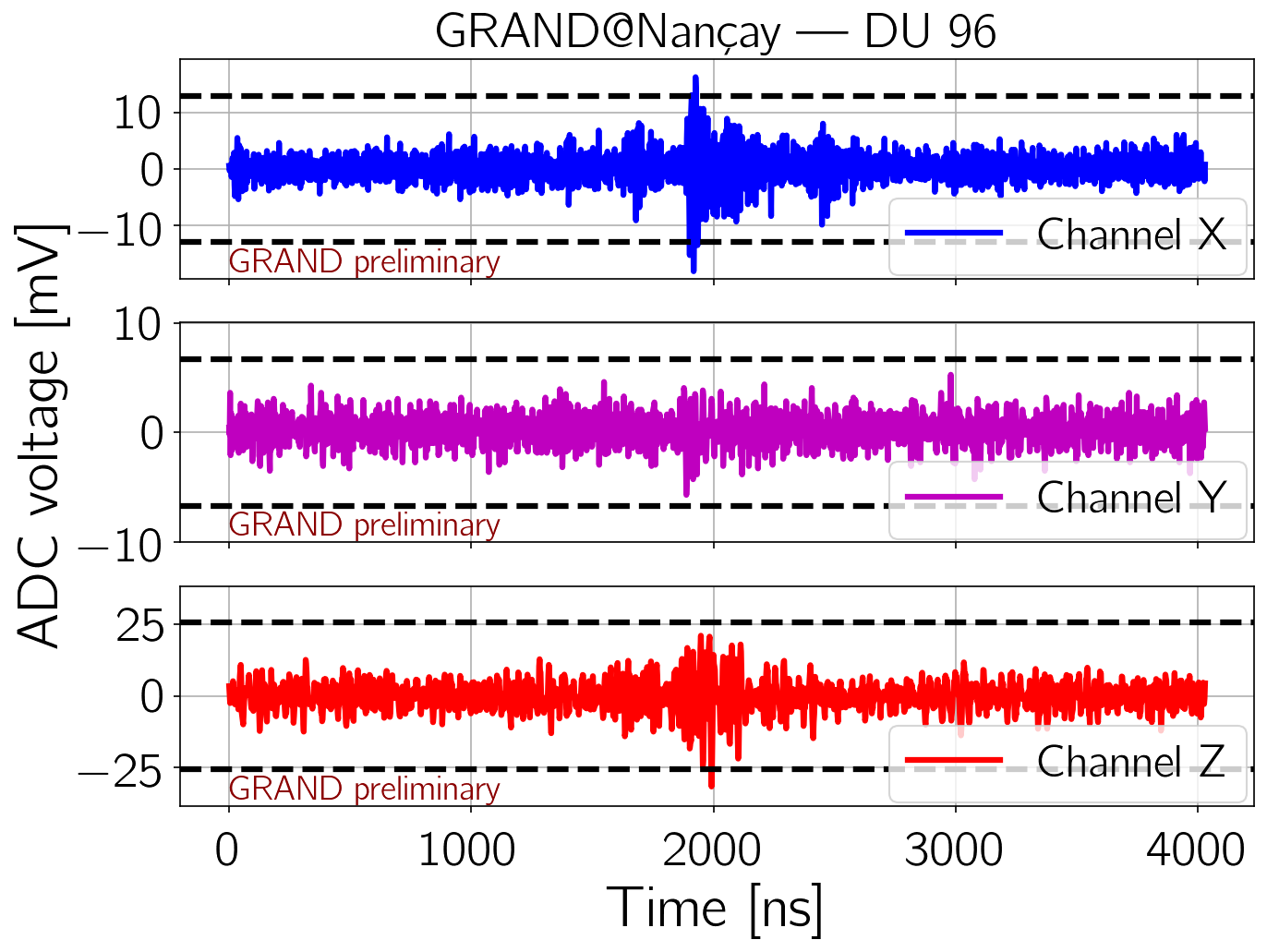}
    \caption{Time trace of a transient RFI event at GRAND@Nançay recorded by DU 96. The top, middle, and bottom panels show the ADC voltages recorded in channels $X$, $Y$, and $Z$, respectively. In each panel, the dashed lines correspond to $\pm5\sigma$, where $\sigma$ is the standard deviation of the trace. In this example, there are six $\pm5\sigma$ crossings for channel $X$, and one $\pm5\sigma$ crossing for channel $Z$. Note that the RFI transient is also slightly visible in channel $Y$, even though it does not cross the $\pm5\sigma$ threshold.}
    \label{fig:trace}
\end{figure}

One of the main quantities relevant to NUTRIG is the ambient rate of transient RFI at GRAND@Nançay, which will be the main background for the FLT. To roughly estimate this rate, we define a transient pulse as follows:
\begin{itemize}
    \item At least one sample of a recorded time trace must exceed $\pm5\sigma$, with $\sigma$ the standard deviation of the trace, in either of the three channels.
    \item If two $5\sigma$ crossings occur in the same trace, but are more than 50 samples (100 ns) apart, they correspond to two different pulses.
\end{itemize}
A typical time trace of a transient RFI pulse recorded at GRAND@Nançay is shown in Fig.~\ref{fig:trace}. It is worth noting that in this example, the pulse is wider than that expected for air-shower signals, which have more short-lived peaked signatures. In our complete data-taking run, which suffers from limited statistics, only a handful of RFI transients were recorded. Given that each recorded trace spans 4032 ns (2016 samples), and that six traces were recorded per minute during our data-taking run (spanning almost 9 hours), we roughly estimate a transient-background rate of the order of several 100 Hz.

\section{Summary and Outlook}
\label{sec:outlook}

The goal of the NUTRIG project is to develop an autonomous online radio trigger for GRAND. This work presented the principle and strategy of NUTRIG, which consists of the development of a first-level trigger (FLT) at the detection-unit (DU) level, a second-level trigger (SLT) at the array level, and a dedicated radio-emission model of very-inclined air showers. A detailed description was given of the GRAND@Nançay prototype, which is dedicated to the NUTRIG project. In particular, it will be used to test the trigger algorithms of the FLT in field conditions.

A brief analysis was performed to characterize the RFI noise at the GRAND@Nançay site. The observed spectrum showed typical features of short waves, FM, aeronautic communications, digital audio broadcasting, and local emitters. In addition, a handful of RFI transients were recorded, defined as pulses that exceed five standard deviations of a time trace. However, only a rough estimation could be made of the overall RFI-pulse rate, which is estimated to be several 100 Hz. 

The next step for the NUTRIG project is to further develop candidate FLT algorithms. These algorithms will first be tested at the LPNHE in Paris, where we already showed that we can feed simulated air-shower pulses to an FEB and recover the pulses in the recorded data. Subsequently, they will be tested in GRAND@Nançay, and further optimized. Doing so, we will also determine which FLT information to provide to the SLT algorithm. The complete FLT+SLT methodology will be tested in the mid-term future at GRANDProto300.

\section*{Acknowledgments}
This work is part of the NUTRIG project, supported by the Agence Nationale de la Recherche (ANR-21-CE31-0025; France) and the Deutsche Forschungsgemeinschaft (DFG; Projektnummer 490843803; Germany). In addition, this work is supported by the CNRS Programme Blanc MITI (2023.1 268448 GRAND; France) and the Programme National des Hautes Energies (PNHE; France)
of CNRS/INSU with INP and IN2P3, co-funded by CEA and CNES. Computations were performed using the resources of the CCIN2P3 Computing Center (Lyon/Villeurbanne, France), a partnership between CNRS/IN2P3 and CEA/DSM/Irfu.

\bibliographystyle{ICRC}
\bibliography{references}

\end{document}